# Methodological Variation in Studying Staff and Student Perceptions of AI

**Juliana Gerard[1*], Morgan Macleod[1], Kelly Norwood[2], Aisling Reid[3], Muskaan Singh[4]**

[1] School of Communication and Media, Ulster University, York St, Belfast BT15 1ED Northern Ireland

[2] School of Psychology, Ulster University, Cromore Rd, Coleraine BT52 1SA, Northern Ireland

[3] School of Arts, English and Languages, Queen's University Belfast, 2 University Square, Belfast BT7 1NN, Northern Ireland

[4] School of Computing, Engineering and Intelligent Systems, Ulster University, Northland Rd, Derry~Londonderry BT48 7JL, Northern Ireland

*** Correspondence:**
Juliana Gerard
j.gerard@ulster.ac.uk

**Abstract**

In this paper, we compare methodological approaches for comparing student and staff perceptions, and ask: **how much do these measures vary across different approaches?** We focus on the case of AI perceptions, which are generally assessed via a single quantitative or qualitative measure, or with a mixed methods approach that compares two distinct data sources — e.g. a quantitative questionnaire with qualitative comments. To compare different approaches, we collect two forms of qualitative data: standalone comments and structured focus groups. We conduct two analyses for each data source: with a sentiment and stance analysis, we measure overall negativity/positivity of the comments and focus group conversations, respectively. Meanwhile, word clouds from the comments and a thematic analysis of the focus groups provide further detail on the content of this qualitative data — particularly the thematic analysis, which includes both similarities and differences between students and staff. **We show that different analyses can produce different results** — for a single data source. This variation stems from the construct being evaluated — an overall measure of positivity/negativity can produce a different picture from more detailed content-based analyses. We discuss the implications of this variation for institutional contexts, and for the comparisons from previous studies.

## 1 Introduction

The integration of generative AI into higher education has accelerated dramatically since 2022, fundamentally disrupting teaching, learning and assessment practices. Universities worldwide deploy tools like ChatGPT for personalised learning, automated feedback, curriculum design and administrative tasks (Salas-Pilco et al., 2023; Zhou et al., 2024). Understanding stakeholder perceptions has, therefore, become paramount, as these perceptions shape adoption trajectories, resistance patterns and the institutional policies governing the educational integration of AI (Chan & Hu, 2023; Zawacki-Richter et al., 2019). Existing research reveals significant ambivalence. Students simultaneously embrace AI's learning potential, while simultaneously fearing academic integrity violations and cognitive dependency (Limna et al., 2022; Rasul, 2023). Staff balance enthusiasm for pedagogical innovation against concerns about erosion of authenticity and professional deskilling (Bearman & Ajjawi, 2023; Lodge, 2024). However, this literature exhibits a key weakness; studies mainly use single methods (typically surveys or interviews) without examining how methodological choices shape rather than merely reveal perceptions.

This blind spot becomes evident when similar contexts yield contradictory findings. Survey-based sentiment analyses routinely report more positive AI attitudes than focus groups within identical institutions (Cotton et al., 2024). Such discrepancies raise fundamental questions; Do variations reflect genuine diversity or methodological artifacts? Without disentangling method from finding, cross-study comparisons become unreliable and policy decisions risk resting on unstable empirical foundations. Despite calls for methodological sophistication in educational technology research (Castañeda & Selwyn, 2018), no study has systematically examined how different qualitative approaches shape findings on AI perception. This gap proves particularly troubling as universities base high-stakes decisions, from integrity policies to pedagogical transformations, on perception studies that may reflect methodological choices more than stable phenomena (Moorhouse et al., 2023). This study addresses this *methodological* lacuna through a systematic comparison of how different data collection and analytical methods construct narratives of AI perception. We shift focus from documenting what stakeholders think to examining how investigative tools shape apparent beliefs. Specifically:

**RQ1:** How do different qualitative data collection methods (questionnaires, digital discussions, focus groups) differentially capture perceptions of generative AI in higher education?

**RQ2:** To what extent do different analytical approaches (sentiment analysis, stance analysis, word clouds, thematic analysis) produce convergent or divergent findings when examining identical phenomena?

**RQ3:** What implications does methodological variation hold for research design and evidence-based AI policy development in higher education?

By emphasising methodology as *constitutive* rather than neutral, this study advances the field in three ways. First, it empirically shows how methodological choices influence AI perception findings, contextualising an expanding body of literature. Second, it explains why triangulation becomes crucial when investigating evolving sociotechnical phenomena. Third, it offers frameworks for designing robust perception studies and interpreting contradictory evidence, ultimately strengthening the empirical basis for institutional AI governance.

# 2 AI perceptions in HE

## 2.1 Single data source

Studies of AI perceptions in higher education usually rely on a single type of data, most often survey responses, interview transcripts, or focus groups. Surveys tend to be analysed through quantitative measures such as sentiment or attitude scales, which allow comparison across populations but flatten nuance (Creswell & Plano Clark, 2011). Interviews and focus groups are typically explored through qualitative thematic coding to capture richer detail and context (Braun & Clarke, 2006). Open-ended survey comments are sometimes coded for polarity or passed through automated sentiment and stance classifiers (Barbieri et al., 2020; Reimers & Gurevych, 2019). Simpler lexical approaches such as word-frequency counts or word clouds can highlight common terms but give little sense of discursive framing (Kucher & Kerren, 2015).

### 2.1.1 Students

Research on student perceptions of AI in higher education generally highlights a predominantly positive attitude, with a strong emphasis on the practical benefits of AI tools. Students frequently report enhanced learning experiences due to increased accessibility and efficiency provided by AI technologies such as generative AI tools (e.g., ChatGPT). These technologies are highly valued for their ability to assist with brainstorming, generating and refining written content, summarizing complex materials, and facilitating research processes (Chan & Hu, 2023; Zhou et al., 2024; Almassaad et al., 2024; Daher & Hussein, 2024). Notably, AI tools are perceived as beneficial in overcoming academic challenges such as writer's block, providing immediate formative feedback, and enabling more personalized, self-directed learning experiences (Chan & Hu, 2023).

Despite these positive attitudes, students also express significant concerns regarding the integration of AI into educational settings. Key apprehensions include issues surrounding academic integrity, particularly regarding plagiarism and originality, as AI tools become increasingly sophisticated and accessible (Zhou et al., 2024). Moreover, students are wary of potential over-dependency on AI, which may negatively affect their critical thinking and problem-solving skills, ultimately compromising long-term educational outcomes and employability skills (Chan & Hu, 2023; Zhou et al., 2024). Consequently, student perceptions reflect a careful balance between the immediate academic advantages provided by AI and deeper concerns over educational authenticity and skill development.

### 2.1.2 Staff

Studies examining staff perceptions of AI demonstrate a nuanced and generally cautious perspective. Staff, encompassing both teaching and administrative roles, recognize the practical advantages of AI tools, particularly in streamlining administrative workflows, enhancing curriculum design, and improving personalized support for students (Salas-Pilco et al., 2023; Zawacki-Richter et al., 2019). AI is also acknowledged as a valuable resource for reducing the time burden associated with routine tasks, allowing educators and support staff to allocate more resources towards strategic planning and direct student engagement (Salas-Pilco et al., 2023).

However, staff perspectives are characterised by notable reservations. Concerns frequently arise regarding the ethical dimensions of AI implementation, particularly about data privacy, algorithmic biases, and the reliability and transparency of AI-generated content (Salas-Pilco et al., 2023). Furthermore, staff often express anxiety about the potential reduction in educational quality and authenticity, emphasising the risk that AI integration could lead to a depersonalised learning experience and compromised student assessment practices (Zawacki-Richter et al., 2019). The apprehension of an over-reliance on AI, which might negatively impact the development of critical academic skills among students, further underlines the cautious approach staff adopt towards widespread AI implementation.

In the current study, we extend the above approaches with a single data source by applying two different analyses to each dataset: for written comments, sentiment analysis and word-level profiles; for focus groups, stance analysis and thematic coding. This enables us to see how conclusions about staff and student perceptions change depending on the analytic lens used, even when the data source remains the same.

### 2.1.3 Students & Staff

When viewed collectively from single data sources, both students and staff acknowledge the significant potential benefits of AI in higher education, although their priorities and concerns differ distinctly. Students primarily value immediate, practical benefits, such as improved efficiency and productivity, directly enhancing their academic performance and learning processes (Chan & Hu, 2023; Zhou et al., 2024). Conversely, staff perspectives frequently emphasize broader institutional considerations, such as the ethical implications, institutional policies, long-term educational outcomes, and the overarching quality and integrity of education (Salas-Pilco et al., 2023; Zawacki-Richter et al., 2019).

The intersection of these perspectives reveals areas of consensus and divergence. Both groups agree on the value AI brings in enhancing productivity and streamlining routine tasks. Nevertheless, the depth and nature of concerns differ, with staff generally demonstrating heightened sensitivity to ethical and long-term implications, whereas students exhibit concerns primarily linked to immediate practicalities and their direct educational experiences (Chan & Hu, 2023; Salas-Pilco et al., 2023). Recognizing and addressing these distinct perspectives is critical for effectively developing AI integration strategies that align with both immediate educational needs and long-term institutional goals.

While these patterns are consistent across the literature, the methodologies used to elicit them vary. Large-scale surveys often foreground headline positives, emphasising efficiency and productivity gains (Chan & Hu, 2023; Zhou et al., 2024). By contrast, interview and focus-group studies elicit more nuanced reservations, particularly staff anxieties around ethics and depersonalisation (Salas-Pilco et al., 2023; Zawacki-Richter et al., 2019). Open-ended survey comments typically reproduce a mix of pragmatic benefits and integrity concerns, but remain less detailed than dialogic data (Cotton et al., 2024). These methodological differences matter: the balance of optimism and scepticism is partly a product of the analytic frame, not only of the underlying attitudes.

## 2.2 Multiple data sources (comparisons)

Studies that combine more than one data source often aim to triangulate findings or to expose contrasts between contexts. Common designs pair surveys with interviews, or open-ended comments with focus groups, in order to check whether headline sentiment aligns with more discursive accounts (Denzin, 1978; Jick, 1979; Greene, Caracelli, & Graham, 1989; Creswell & Plano Clark, 2011). Quantitative survey data are usually subjected to descriptive statistics or sentiment scales, while qualitative transcripts are thematically coded to surface patterns of concern and opportunity (Braun & Clarke, 2006). Some more recent studies incorporate lexical profiling or stance analysis to compare tone and alignment across sources (Vu et al., 2025; Cotton et al., 2024).

### 2.2.1 Students

When comparing student perceptions of AI across multiple qualitative data sources, certain insights emerge that are less evident in single-method studies. For instance, surveys habitually highlight predominantly positive attitudes related to efficiency and ease of use, whereas focus groups tend to elicit more reflective and critical discussions about ethical considerations and dependency on technology (Chan & Hu, 2023; Vu et al., 2025). Structured comments, such as those obtained through platforms like Padlet, frequently use AI tools in everyday academic tasks. However, during interactive focus groups, students typically reveal deeper concerns related to the loss of personal interaction, reduced opportunities for critical thinking, and the ethical implications surrounding AI-generated content and originality (Vu et al., 2025; Zhou et al., 2024). This variability highlights the importance of methodological triangulation in understanding complexity of students' perceptions about AI in education.

### 2.2.2 Staff

Staff perceptions similarly benefit from analysis across multiple qualitative methodologies, revealing layers of complexity not fully captured by single-source studies. For instance, while written feedback via surveys often emphasises practical efficiencies such as administrative simplification and teaching support, more structured interactions through platforms like Padlet or intensive dialogues in focus groups highlight significant ethical and pedagogical concerns (Salas-Pilco et al., 2023; Vu et al., 2025). In focus groups, staff tend to express heightened concern regarding AI's potential impact on pedagogical quality, assessment authenticity, and the professional roles of educators, intimating deeper underlying anxieties about institutional changes and job security (Zawacki-Richter et al., 2019). The interaction between different qualitative methodologies illustrates how the depth and specificity of staff concerns can vary significantly, depending on the context and medium of data collection.

### 2.2.3 Students & Staff

Comparative analyses using multiple qualitative methods and integrating the perspectives of both students and staff provide richer, more complex insights into perceptions of AI. Participants across groups widely acknowledge practical benefits related to AI, especially its capacity to enhance efficiency and productivity. However, notable differences emerge regarding concerns around dependency, ethical use and the broader educational implications of AI (Chan & Hu, 2023; Salas-Pilco et al., 2023; Zhou et al., 2024; Arowosegbe et al., 2024; Yan et al., 2025). Students typically maintain a pragmatic stance, focusing primarily on immediate academic advantages, with their structured comments and focus group discussions frequently emphasizing task-specific benefits. In contrast, staff responses across various qualitative methods increasingly highlight systemic issues, including changes in pedagogical practices, academic integrity challenges, and broader institutional policy considerations (Vu et al., 2025; Zawacki-Richter et al., 2019).

The intersection of diverse qualitative methodologies thus enriches understanding by highlighting not only the common ground between stakeholder groups but also distinct focal points and depths of concern. Integrating multiple qualitative data sources ultimately supports a more holistic approach to policy development and strategic planning for AI integration in higher education, acknowledging both practical day-to-day implications and overarching institutional impacts.

The current study extends this multi-source tradition by directly comparing three forms of qualitative data (i.e. free-form comments, structured Padlet discussions, and semi-structured focus groups) with four analytic lenses; these include sentiment analysis, stance analysis, word-level profiles and thematic coding. This allows us to evaluate not only whether staff and student perceptions converge but also how far that convergence depends on the methodological combination chosen.

## 2.3 Summary and current study

Numerous existing studies have examined perceptions of artificial intelligence in higher education using various quantitative and qualitative methods. Typically, quantitative approaches utilise structured surveys or scales, enabling researchers to quantify levels of acceptance, usage frequency, and general attitudes toward AI integration. Qualitative approaches, on the other hand, include open-ended surveys, structured interviews, focus groups, and digital discussion boards, providing richer, narrative-based insights into participants' perceptions, concerns, and experiences (Salas-Pilco et al., 2023; Zawacki-Richter et al., 2019). Despite the availability of diverse methodologies, there remains limited research systematically comparing insights derived from multiple qualitative approaches. Most qualitative studies rely heavily on a single data collection method, such as individual written comments or group interviews, without evaluating whether different qualitative methods yield consistent or varying perceptions.

To address this methodological gap, the current study explicitly compares perceptions collected from multiple qualitative data sources: free-form written comments from a survey, structured written comments posted publicly on Padlet, and semi-structured focus group interviews. We then present four analytic approaches to comparing AI perceptions across students and staff: a sentiment analysis and a stance analysis, both of which compare the extent to which staff and students express positive, negative or neutral sentiments towards AI in education contexts; a word cloud analysis, which identifies frequent words used across the different populations; and a thematic analysis, which identifies key themes that appear in the data across the different populations.

Crucially, the different analytic approaches highlight different considerations. With this comparative approach, we aim to identify the consistency and variation in AI perceptions across different qualitative methodologies and among distinct stakeholder groups within higher education, including students, teaching staff, and non-teaching staff. By examining how these perceptions differ or align based on the qualitative method employed, this study seeks to inform future AI policy, practice and research within educational institutions.

## 3 Materials and methods

In this section, we outline three approaches for investigating perceptions AI in education settings, all using qualitative data: free-form written comments as part of a broader questionnaire, structured written comments and replies on a publicly available Padlet, and focus groups. For all three methods, the sample was recruited from staff and students at Ulster University in Northern Ireland.

### 3.1 Participants

Free-form written comments were provided by a total of 292 staff and students. These populations were further delineated for students into undergraduate/postgraduate taught (n=185) and postgraduate research (n=26), and for staff as those with teaching responsibilities (n=31), and non-teaching staff — including Professional Services, Administrative and Management (n=50). Demographic breakdowns for these participants are provided in Tables 1 (age range) and 2 (gender).

**Table 1. Participant demographics for free-text comments: age range**

| | Age range | | | | |
|---|---|---|---|---|---|
| **Population** | **18–24** | **25–39** | **40–59** | **60+** | **Not provided** |
| Undergraduate & Postgraduate taught | 80 | 79 | 25 | 1 | 0 |
| Postgraduate research | 5 | 12 | 9 | 0 | 0 |
| Teaching staff | 0 | 11 | 14 | 5 | 1 |
| Non-teaching staff | 1 | 19 | 25 | 5 | 0 |

**Table 2. Participant demographics for free-text comments: gender**

| | Gender | | | | |
|---|---|---|---|---|---|
| **Population** | **Female** | **Male** | **Non-binary** | **Other** | **Not Provided** |
| Undergraduate & Postgraduate taught | 96 | 81 | 5 | 2 | 1 |
| Postgraduate research | 15 | 11 | 0 | 0 | 0 |
| Teaching staff | 11 | 18 | 0 | 1 | 1 |
| Non-teaching staff | 34 | 16 | 0 | 0 | 0 |

Structured written comments to the Padlet were collected from the broader categories of staff (n=54) and students (n=124); further demographic information was not collected.

Finally, the focus groups included 11 participants in total; three separate focus groups were comprised of undergraduate/postgraduate taught and a researcher (all with no teaching responsibilities) (n=4, 3 females and 1 male), staff with teaching responsibilities (n=3, 2 females and 1 male), and non-teaching staff (n=4; 2 males)[1].

All three approaches received ethical approval from Ulster University, and participants received a £10 Amazon voucher for their participation in the questionnaire containing the free-form text comment, and in the focus groups. Participants for the questionnaire and the Padlet were recruited via university-wide email listservs to staff and students, while the focus group participants were recruited from the sample who had completed the questionnaires.

### 3.1.1 Free-form text comments: perceptions questionnaire

The free-form text comments were collected as part of a broader questionnaire on AI awareness and perceptions, to investigate changes in perceptions over time (in preparation). In the questionnaire, participants responded first to a series of questions with Likert scale (i.e. quantitative) ratings about awareness and perceptions, followed by demographic questions. Finally, participants had the option to provide further details on their perceptions in the form of the free-text comment, with the following prompt:

*(Optional) "Please enter any further comments about this questionnaire, your experience with AI in academia, or anything else that you would like us to consider."*

We focus on responses to this prompt. As this field was optional, not all respondents included a comment; the current analysis therefore does not include comments from additional participants who left this field blank (n=456), who entered a comment like "n/a," or who commented on the questionnaire itself rather than their perceptions of AI (e.g. "I like this questionnaire") (n=11).

[1] Focus groups were selected over interviews in order to generate discussion between group members.

## 3.2 Design

### 3.2.1 Focus groups

The focus groups were designed to address key topics that have been included in previous quantitative surveys (e.g. Petricini et al., 2024), including:

- experiences of AI
- perceptions of AI in teaching and education
- opportunities
- concerns and risks of AI and education
- facilitating future use of AI

A semi-structured interview guide was utilized to explore these topics, which allowed for follow up and elaboration depending on participants' responses. The three focus groups (one each with students, teaching staff and non-teaching staff) were held and transcribed using Microsoft Teams and ranged from 41-55 minutes in length.

### 3.2.2 Structured text comments: Padlet

Padlet comments were collected from staff and students based on the protocol developed by Drumm et al (2023): in contrast with the free text questionnaire comments, Padlet allows for anonymous posting *and* commenting on posts, meaning that many of the Padlet comments were provided as responses. Students and staff posted to separate Padlets, but were both presented with the following prompts for posting:

- Have you used any of Generative AI tools? If yes, which ones?
- How have you used these tools?
- If you are not currently using any of these tools do you plan to use them in the future?
- If your answer is yes, why do you intend to use them?
- If your answer is no, why do you intend not to use them?
- How do you think others are using these tools?
- What do you think of others using generative AI?
- Should Generative AI play a role in learning and assessment?

We analysed Padlet data of qualitative responses (160 entries post-cleaning: 114 students, 14 staff). Each entry was labelled as by student or staff. The Padlet format allows users to reply to each other's comments in threads. Replies were excluded from analysis if their content was too closely tied to that of a previous post (e.g. 'I agree'), but were included as separate cases if they had enough independent content of their own. As with the survey data, irrelevant responses were also excluded.

## 3.3 Analysis plan

To analyse the data from the above approaches, we focus first on the free-text and Padlet comments, and second on the focus groups. We deploy two separate analyses for both datasets — respectively, the comments and focus groups — allowing for a direct comparison between different analyses with the same dataset. For both datasets, the first of these analyses provides a more surface-level understanding of the data, collapsing over the content (sentiment analysis and stance analysis); in contrast, the second analysis offers further insight on the content (Word cloud, thematic analysis).

#### 3.3.1 Free-text and Padlet comments

##### 3.3.1.1 Sentiment analysis

One aspect of the analysis performed here involved quantitative analysis of sentiment scores. This analysis made use of automated sentiment scores provided by the RoBERTa-based Cardiff TweetEval model (Barbieri et al., 2020), a model that was also used in Gerard et al. (2025) with a different dataset. Automated sentiment scores were generated for student and staff comments from the initial survey, as well as scores from the Ulster Padlet data. The automated scores were then compared with manually assigned sentiment scores, in order to evaluate the effectiveness of such automated analysis.

##### 3.3.1.2 Wordcloud analysis

Next, we extracted frequently used words from the free-text and Padlet comments to generate word clouds for each population group. This analysis provides greater insight on the content of the comments — albeit at the word level without further sentence context. Importantly however, differences in frequent word choice can be indicative of variation in the specific content of the comments across populations.

#### 3.3.2 Focus groups

##### 3.3.2.1 Stance analysis

For the focus groups, we first incorporated a *stance and alignment* perspective, following Du Bois's (2007) *stance triangle* model (Park et al., 2024). In this view, a speaker takes a stance by evaluating an object (e.g. AI), positioning themselves, and *aligning or disaligning* with other speakers' stance, which allows us to analyse how participants responses positioned themselves relative to AI (stance). Like the sentiment analysis, this produced positive, neutral and negative tags, but for the focus group utterances rather than individual comments. This allowed for a quantitative analysis of the stance tags across the population groups. For stance detection, we utilized eevvgg/StanceBERTa,[2] a fine-tuned DistilRoBERTa model trained on socio-political texts, which classifies each utterance as expressing support, opposition, or neutrality toward a given topic.

Sentiment analysis was carried out using cardiffnlp/twitter-roberta-base-sentiment,[3] a model fine-tuned on over 124 million tweets via the TweetEval benchmark, particularly well-suited for informal and short-form text such as focus group speech. We applied browndw/docusco-bert,[4] a model trained on the DocuScope corpus, which assigns labels such as contrast, support, and examples to capture discourse roles and argumentative framing; these annotations allowed us to layer thematic clusters with discourse-level metadata, producing a multidimensional view of each utterance that includes its semantic topic, stance, emotional tone, rhetorical function, and dialogue role.

##### 3.3.2.2 Thematic analysis and NLP themes comparison

Focus group data were analysed using inductive, thematic analysis using a six-phase process: data familiarization, generating initial codes, searching for themes, reviewing the themes, defining and naming the themes and report writing (Braun & Clarke, 2006). Transcripts were read twice for familiarization, then generated initial codes. All transcripts were reanalysed using the final set of agreed codes reflecting the iterative process.

[2] https://huggingface.co/eevvgg/StanceBERTa

[3] https://huggingface.co/cardiffnlp/twitter-roberta-base-sentiment

[4] https://huggingface.co/browndw/docusco-bert

Additionally, we used Python-based NLP libraries for text processing (e.g., NLTK, spaCy) and data analysis to compare the themes across populations, providing further insight into variation across the population groups. To do so, we implemented a modular NLP pipeline combining pre-trained transformer-based models and unsupervised clustering techniques. Each utterance was embedded using the all-MiniLM-L6-v2 SentenceTransformer,[5] which provides compact, high-quality semantic embeddings trained on natural language inference (NLI) and semantic textual similarity (STS) tasks.

# 4 Results

## 4.1 Comments

### 4.1.1 Sentiment scores

The mean sentiment scores for students and staff are presented in Figure 1, for both the free-form text comments in the questionnaire and the Padlet comments (numeric scores are provided in the Appendix). As described in Section 3.3.1, the RoBERTa model produced scores on a 3 point scale, designating comments as positive (1), neutral (0), or negative (-1). To facilitate comparison, manual sentiment scores were expressed in the same format; Figure 1 includes both scoring formats (RoBERTa and manual).

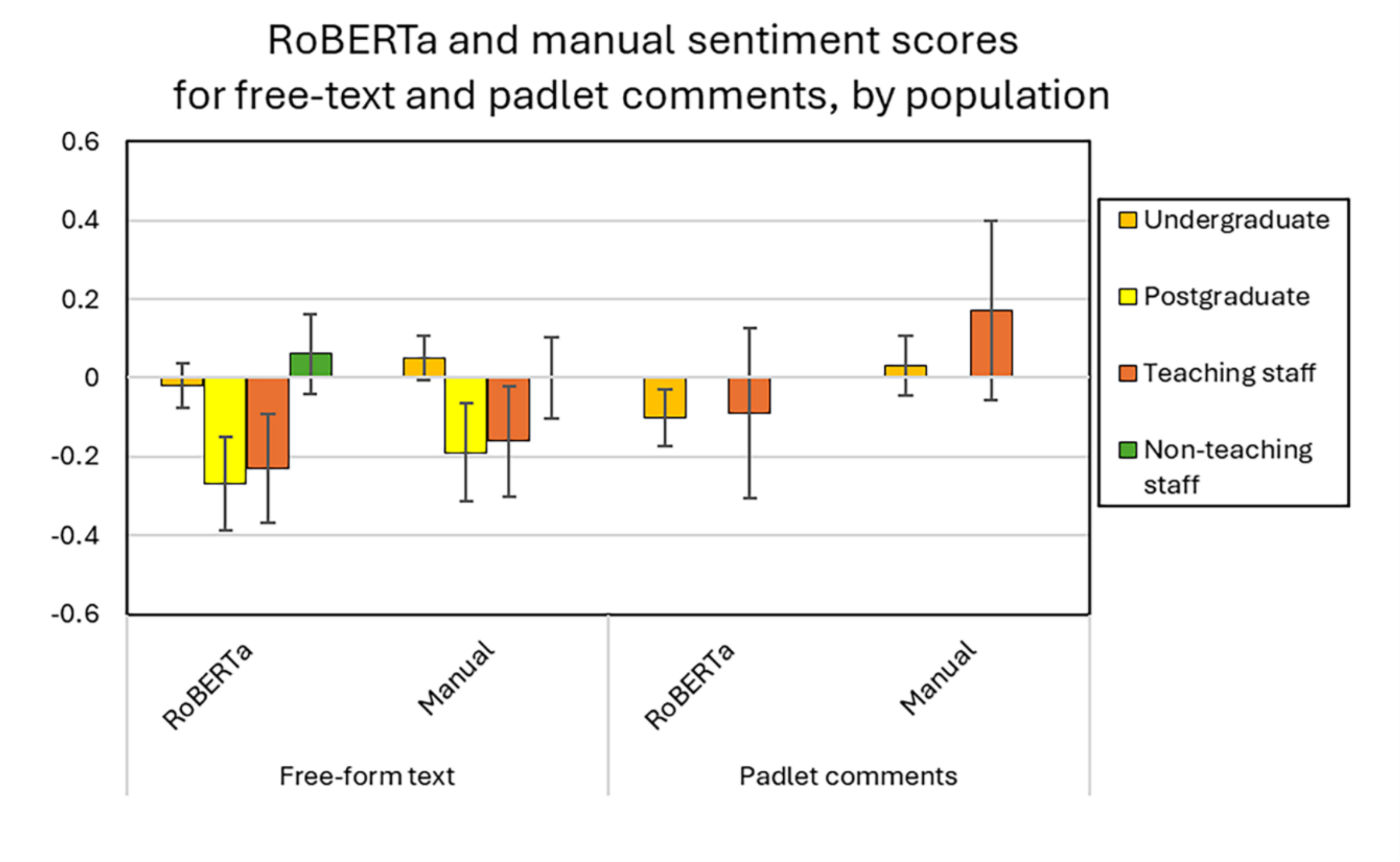

**Figure 1. Mean sentiment scores - both RoBERTa model and manual scoring on -1 to +1 scale.**

[5] https://huggingface.co/sentence-transformers/all-MiniLM-L6-v2

#### 4.1.1.1 Sentiment score analysis

To compare the sentiment scores in Figure 1 across populations, separate linear regressions were conducted for the free-form text comments and Padlet comments, respectively (due to the different levels for the *population* factor). We also compared the RoBERTa and manual scores, to assess reliability across the two scoring methods. Accordingly, for both analyses — free-form text comments and Padlet comments — the dependent measure was the sentiment score, with fixed effects as population (e.g. staff/students) and scoring method (automated/manual).

First, for the free text comments, there was a significant main effect of population ($F$ = 3.002, $p$ = 0.03), indicating a significant difference between one or more specific population groups in Figure 1, for the free-text comments. To identify which groups, we conducted pairwise post hoc comparisons between the different population groups, presented in Table 3.

**Table 3: Pairwise post-hoc comparisons - free text**

| Pairwise comparison | F | p |
|---|---|---|
| Undergraduates — postgraduates | 0.247 | 0.108 |
| Undergraduates — teaching staff | 0.210 | 0.163 |
| Undergraduates — non-teaching staff | -0.014 | 0.998 |
| Postgraduates — teaching staff | -0.037 | 0.993 |
| Postgraduates — non-teaching staff | -0.260 | 0.165 |
| Teaching staff — non-teaching staff | -0.224 | 0.239 |

Surprisingly, while the overall effect of population was significant, the pairwise post hoc comparisons revealed that none of the individual pairwise comparisons were significant, i.e. the numerical differences between population groups in Figure 1 were *not* in fact significant for the free-text comments. Moreover, the effect of population was not significant for the Padlet regression model ($\beta$ = -0.148, $Z$ = 0.14, $p$ = 0.293). Thus, while numerical differences can be observed in Figure 1 between the different population groups, these differences were not significant.

Additionally, there was there was no difference in automated RoBERTa scores vs manual scores - for neither the free text comments ($F$ = 0.236, $p$ = 0.627) nor the Padlet comments ($\beta$ = -0.261, $Z$ = 0.167, $p$ = 0.12), and no interaction between population and scoring method — for neither the free text comments ($F$ = 0.231, $p$ = 0.875) nor the Padlet comments ($\beta$ = 0.138, $Z$ = 0.198, $p$ = 0.486). Together, these results confirm that there were no significant differences between the population groups for *either* scoring method.

Thus, from a quantitative perspective of positive, neutral and negative sentiments, students and staff appear similar in their attitudes toward AI — a surprising result, given the previously observed differences outlined in Section 2 (e.g. Chan & Hu, 2023; Salas-Pilco et al., 2023). Next, to better understand this result, we assess the reliability of the sentiment score measure.

#### 4.1.1.2 Reliability check: automated and manual scoring

As observed in the previous section, there was broad agreement between the automatic and manual scores, with no significant differences observed for free-text or Padlet comments. In particular, the automated and manual scores were found to correspond in 79% of cases (356/452). In this section, we explore the source of the remaining divergences.

Some divergences were the result of borderline cases, comments containing a mixture of positive and negative statements where legitimate disagreement would be possible as to which, if any, predominated. For the purposes of this comparison, these were not scored as mismatches. Of the remaining cases, 20% (91) represented a difference of one point in the LLM's three-point scale, i.e. scoring a positive or negative comment as neutral, or vice versa. Only 1% (5) represented a greater difference; all 5 of these involved a comment that was scored as negative with automated scoring, but positive with manual scoring. This accuracy rate can be expressed in the form of a confusion matrix, as in Table 4.

**Table 4: Confusion matrix**

| | **Manual Score** | | |
|---|---|---|---|
| **Automated Score** | Negative | Neutral | Positive |
| Negative | 112 | 25 | 5 |
| Neutral | 18 | 145 | 36 |
| Positive | 0 | 12 | 101 |

(1) *Ai helps to explain in detail where I have confusion in my academic work.* (automated score: neutral; manual score: positive)

(2) *I have limited experience with AII have spoken to others with more knowledge in the area I think I could be helpful but I do not know where or how the information is gathered for AI use* (automated score: negative; manual score: positive)

(3) *This questionnaire focused on AI text-generators. I am not a fan of these because I believe the technology has shaky foundations ethically speaking, and I also feel they may lead to poor quality resources from instructors or poor quality work from students. I feel the same way about AI image generation. However I do think there is a lot of potential for AI in other fields - I would be interested in learning how to code using AI, and as a Geography student I think there is potential for some very interesting developments in mapping or modelling software using this technology.*
(automated score: negative; manual score: neutral)

In ((1)), the neutral score given by the automated analysis may result from the relative lack of overt emotive language in the comment, or possibly from the combination of positive *helps* with negative *confusion*. In ((1)), the LLM has correctly recognized that the comment reflects negatively on the respondent's lack of knowledge; while the attitudes expressed towards AI are more positive than negative (*could be helpful*), the model is simply providing a general sentiment score, as designed, rather than anything related to a specific topic such as AI.

An example such as ((2)) illustrates the limitations of any purely quantitative approach to sentiment analysis, whether automatic or manual. In this example, the respondent provides a nuanced discussion of the pros and cons of multiple forms of AI. A numerical score merely reflects an attempt to balance the positive and negative components of such a comment and to decide which, if any, predominated. However, the individual components of such a response are worth consideration separately, in their own right.

### 4.1.2 Word clouds

To analyse the individual components of staff and student comments, we explored the different words used by students and staff in their comments across the two datasets. If the similarities observed in overall sentiments between students and staff are indeed indicative of similar perceptions, then we expect to observe similar words used their comments about these perceptions. However, if their perceptions are more varied across populations, despite the similarities observed for overall sentiment scores, then we expect this variation to be reflected in their word use.

To test this prediction, we generated word clouds for staff and student comments across the free text and Padlet comments.[6] The word-cloud generator itself is capable of performing some screening for common stop words (e.g. *and*, *the*). However, a manual review was conducted for additional screening, to remove words with little semantic content (e.g. *actually*) and words that reflect constant aspects of the study design (e.g. *university*, *AI*). The resulting word clouds are presented in Figure 2. Key similarities across populations stand out in Figure 2, consistent with the similarities across sentiment scores. In particular, the top 2 words across both populations, for both data sources, are "tool" and "work." The frequent use of these terms is indicative of common themes around AI use in education contexts, e.g. perceiving AI as a tool for a range of applications.

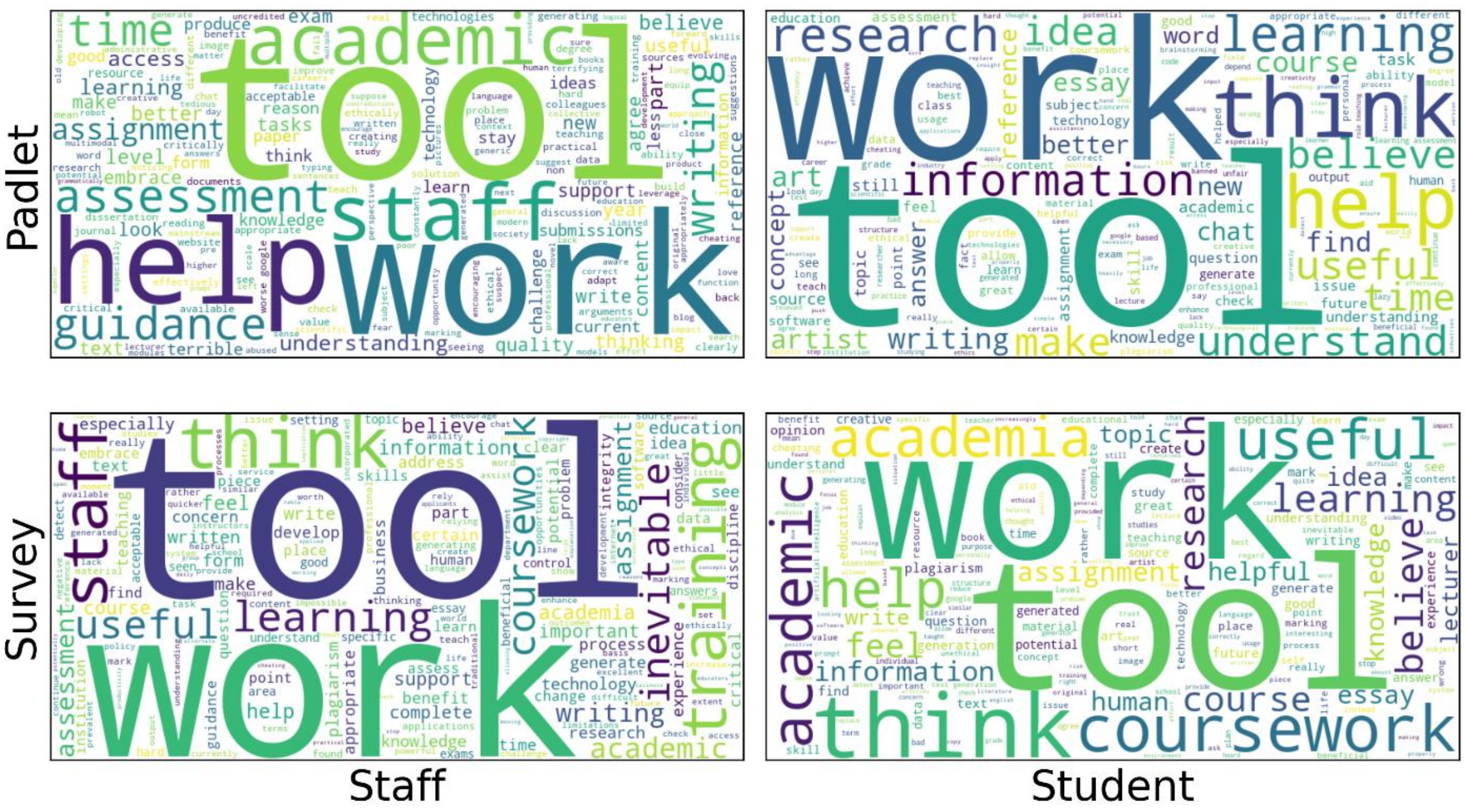

**Figure 2. Word clouds for students and staff in their free text comments (Survey) and Padlet comments.**

However, some variation is also observed between staff and students, across both data sources, when it comes to other frequently used words. For example, the next most frequently used words in free-text comments were "coursework" for students, but "training" for staff. This contrast may reflect different priorities — students are concerned about AI use in coursework, while staff are concerned about adequate training; however, previous research has observed these priorities across both populations. Moreover, a frequent word for both students and staff in the Padlet comments was "help"— which may relate to any of the above. Further exploration of this variation requires a more in-depth analysis of the content itself. We address this in the following section, with a thematic analysis.

[6] https://github.com/amueller/word_cloud

In sum, the word cloud analysis reveals some variation in the content of comments from staff and students. This contrasts with the sentiment analysis, which did not find significant differences between these population groups. That is, different measures with the same dataset produced a different result. In the following section, we expand this approach to the focus groups, using a Stance analysis to identify negative, neutral and positive utterances, and a thematic analysis to explore the content of staff and student perceptions in more detail.

## 4.2 Focus groups

### 4.2.1 Stance analysis

Stance was compared across staff and student utterances, by categorizing each utterance as "support," "oppose," or "neutral." In contrast with the sentiment analysis presented in Section 4.1 — which focused on standalone comments — these categories were differently distributed across staff and student utterances in the focus group (Figure 3).

Notably in Figure 3, a higher proportion of staff comments tended to skew more negative or neutral. Very few staff utterances were explicitly supportive of AI's role (many were neutral descriptions or mixed), and a notable subset were actively critical (e.g. calling AI output *"pointless"* or lamenting its misuse). Students, by contrast, had a higher proportion of supportive sentences (e.g. praising AI's helpfulness). Even so, the majority of both student and staff remarks were tagged Neutral in stance — often simply describing experiences or making factual statements without overt judgment.

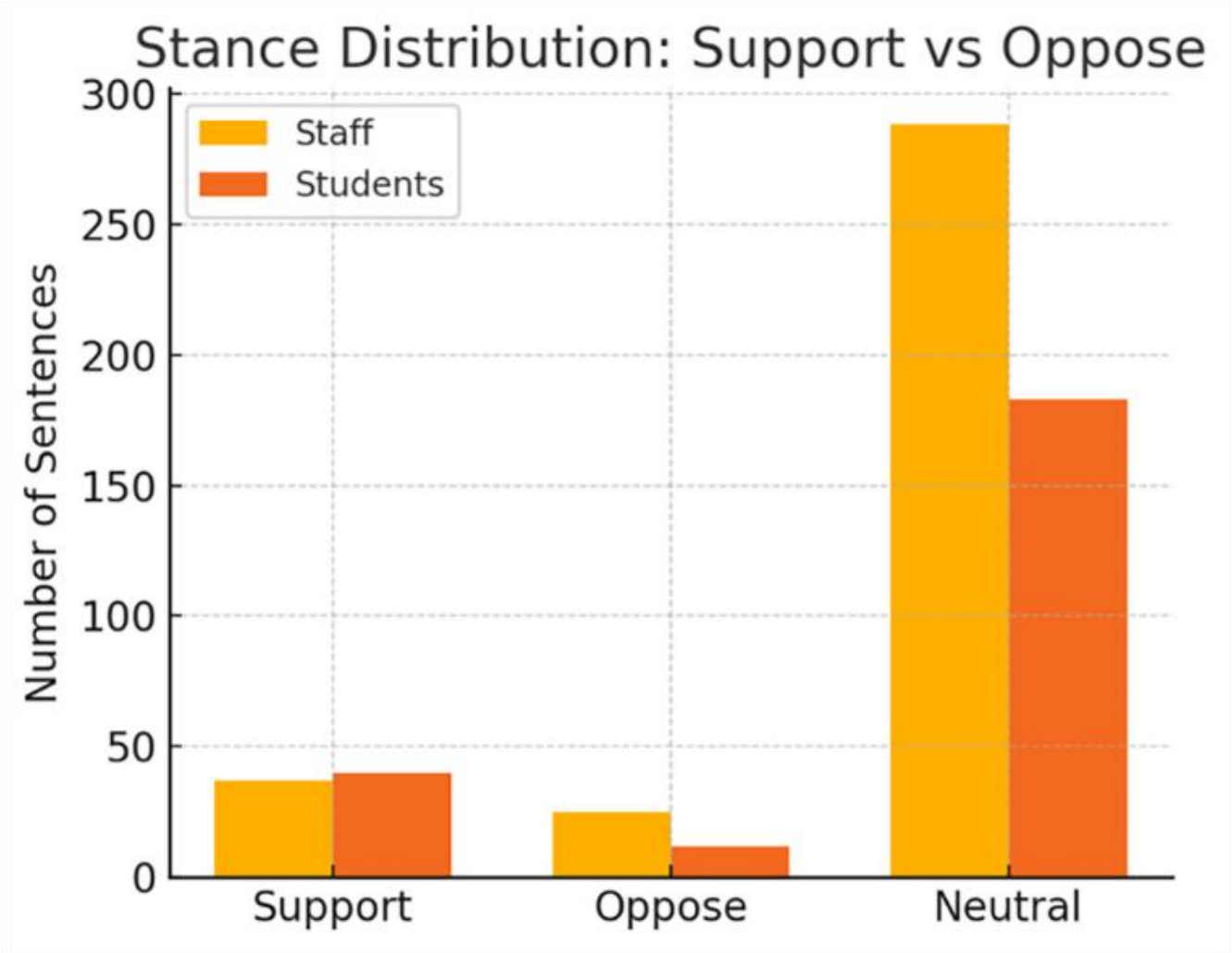

**Figure 3: Stance Distribution across student and staff participant statements.**

### 4.2.2 Thematic analysis

Next, the focus groups provided an opportunity for more in-depth exploration of students' and staff perceptions surrounding the same general topics relating to AI in education contexts, via thematic analysis. Main themes and sub-themes from the focus groups were identified as described in Section 3.3, and are presented in Table 5.

**Table 5. Focus group themes and sub-themes**

| Theme | Sub-theme |
|---|---|
| Attitudes to AI | - Prior experience with AI<br>- Positive Attitudes towards AI<br>- Negative Attitudes towards AI |
| Highlights & Challenges of AI application in HE | - Opportunities that develop student learning<br>- Obstacles to effective student learning<br>- Optimizing AI to enhance education |
| The future of AI | - Staff and student training |

#### 4.2.2.1 Theme 1: Attitudes to AI

Attitudes appeared to be affected by prior experience, as the more proficient users discussed the use, and growth, of AI in academia positively, with ChatGPT being the most frequently used AI platform — consistent with previous quantitative studies (Amoozadeh et al., 2024; e.g. Faruk et al., 2023). Participants highlighted the utility of AI in supporting administrative tasks, research activities and the development of teaching materials. Notably, they also discussed AI's current limitations, particularly the necessity for fact-checking and the importance of crafting effective prompts to generate tailored and accurate outputs. Negative attitudes were often observed in response to others' use of AI. For example, some noted instances where AI-generated text included overly complex vocabulary, irrelevant content and poorly structured sentences, which suggested a lack of user engagement with fact-checking or editing content.

Key differences emerged across the focus groups: support staff (SS) and non-teaching students and researchers (NTS) had significant prior experiences and held positive attitudes. Comparatively, teaching staff (TS), were mixed in terms of their prior experience and expressed similar amounts of positive and negative attitudes, e.g. *'they're [students] becoming lazier'* (TS). A contrast emerged between the TS and NTS in their attitudes towards students' capabilities when using AI, as TS felt that students were not aware of the limitations (mentioned above) but most of the NTS expressed their own awareness of the limitations and discussed how they managed these within their work. Interestingly, when discussing their prior experiences, no one reported examples of AI being embedded within teaching practices.

#### 4.2.2.1.1 Sub-theme: Prior experience with AI

Commonalities in AI usage were evident across groups, for example, when structuring work, 'getting started' when deadlines were close, and not starting with a 'blank slate'; common uses also included generating and editing images:

> *'If you need to remove a part of an image, it's quite good at being able to analyze the image and take something out…that you don't need anymore' (SS1),*

NTS and SS found AI models to be effective search engines and more enhanced than Google, as they were capable of synthesizing large amounts of information. Further, AI was effective for some research activities, such as 'tight turnaround' research funding applications.

TS deemed AI useful for teaching when developing tutorial questions, preparing lectures, and planning induction sessions. These were all described as a means of generating and refining ideas before fully developing them into comprehensive materials:

> *"I need an idea. Let's flush it out and then I pick up the ball and I run with it, and I might ask it for bullet points" (TS).*

For the NTS, AI was used academically to reduce word count, summarize extensive information, understanding new or complex concepts and definitions, the preparation of professional documents, such as CVs and cover letters, and when selecting dissertation topics. From the SS, one participant described themselves as an 'early adopter' and was very comfortable and skilled with using AI models. SS discussed AI for administrative tasks and developing PowerPoint presentations:

> *"A lot of times you're having to do research at the drop of a hat about organizations or whatever it might be to draft press releases. And I've found AI a really good kind of research too, in that, and it's definitely got better" (SS).*

Relating to research, NTS (and to a lesser extent the TS group), discussed the use of AI when structuring and drawing out key points for the introduction sections of publications. Specific to NTS, AI was beneficial when trying to understand others published research and a means of simplifying and explaining material in more accessible, lay terms:

> *"I would use ChatGPT quite regularly, probably a few times a week….if there's like a topic that I'm trying to read up on, and it's got loads of new concepts in it and I'm like looking at sources online that I don't really understand then I will ask ChatGPT to understand it a bit better" (NTS).*

#### 4.2.2.1.2 Sub-theme: Positive attitudes towards AI

All groups expressed positive attitudes relating to AI helping to start work, with some acknowledging that this was especially helpful for the neurodivergent population:

> *"In particular like for those…with say, for example ADHD, it's getting started and it's fine when they get started" (SS).*

AI was also commended in teaching and learning practices particularly as a means of initiating the process and reducing administrative tasks:

> *"If you need a course philosophy rather than designing your own course philosophy, you can ask ChatGPT to write one for you" (TS).*

All groups were positive in the use of AI for image generation and editing as it is user friendly, with a further expressed benefit of avoiding copyright issues:

> *"I usually get a ChatGPT image which is generally copyright free…. I think joining images together is quite interesting and when moving away from text" (TS).*

All groups felt that AI was 'faster than humans' and was especially helpful when time was limited as a search engine, and for synthesizing large amounts of information. Further, the NTS and SS groups discussed the benefits of AI as a personalized learning tool and mentoring pathway — which is further developed under T2 (Highlights & Challenges of AI application in HE):

*"It's really good for that ideation side of things like getting the ideas for what the direction you want to go in and then acts as like that research tool afterwards, you did still have to do a lot of fact checking, but it was amazing to see what Chat GPT could come up with" (SS1).*

Participants expressed that AI was a good learning tool for the student population, when they learned how to develop effective prompts, therefore, with increased usage, users learnt to develop effective prompts leading to tailored output specific to their learning needs and discipline. Positive attitudes were evident when discussing the limitations associated with AI and how they negated these to extract meaningful content. The more proficient users outlined the need to develop good prompts, 'tweaking the output', fact checking and screening out spurious content. No positives were reported by TS, when discussing students using AI to aid coursework production. Further, knowledge exchange was evident during the discissions as participants described their practices, e.g. 'I didn't think it could do something like that, so that's good' (NTS).

#### 4.2.2.1.3 Sub-theme: Negative attitudes towards AI

Most participants were aware that AI models can produce spurious and incorrect content, which was seen as a problem:

*"Some of the stuff we get back is, well, pure garbage, but at least we have the skill set to filter that out" (TS).*

Concerns were voiced relating to younger generations being 'impressionable' and believing the incorrect content to be true. Several participants stated that it was quite frightening. An example was given whereby one participant from the NTS group identified AI content in a group presentation due to the content being irrelevant to the topic:

*"I think it can be obvious sometimes when people are using it and but…the kind of scary part is people just relying on that and not like checking what's coming out and then just kind of go on with whatever Chat GPT says" (NTS)*

TS expressed negative attitudes regarding students' use of AI and the 'dumbing down' and 'cheating' in student learning. For example, some noted that students are increasingly using AI in their coursework which they felt was easy to identify in the 'verbiage and overtechnical writing' and were disappointed to see that students were using ChatGPT to compose basic forms of communication, such as emails. Further, they also recognized AI output in students' coursework, undergraduate applications and personal statements, and in response to calls for external examiners, which was accompanied by a sense of 'depersonalization'.

Depersonalization was discussed within the TS and SS groups, for example, when AI was used to complete applications from staff and students external to the university:

*'They just get chat GPT to write their application and again it doesn't give me any sense of what they are' (TS).*

This was echoed within the SS group as AI does not 'have a heart or feelings' and when individuals use AI to create presentations for job interviews their sense of self and passions are diminished. This aligned with discussions focused on 'social reduction' and the worry that students may become more withdrawn and replace human interactions with AI, and then they 'become more like robots'.

Outside of student learning, concerns were voiced by the SS group in relation to opinions on workloads as individuals do become more proficient AI users. For example, if AI was used effectively to complete administrative tasks, workload may increase over the longer-term and specialized roles could become more general, i.e. '*The more we embrace it, the more roles might be muddied'* (SS).

#### 4.2.2.2 Theme 2: Highlights & Challenges of AI application in HE

Participants reflected on the dual potential of AI to both enhance and hinder university education. They explored strategies for maximizing AI's benefits and reducing its limitations and gave insights as to how AI could be best utilized within the university context. The SS and NTS groups viewed AI as providing opportunities rather than obstacles. They emphasized the potential for AI to enhance learning experiences, while acknowledging concerns related to responsible and ethical usage, such as transparency in citing AI use — topics further developed under Theme 3 (The future of AI). Opportunities to further develop student learning are personalized learning and mentoring, accessibility and support. The NTS group showed a thoughtful approach to AI — they recognized its potential to support learning, but some expressed strong views against its use in assessments, particularly automated marking. The NTS group discussed AI's uses but had not used AI models to obtain personal feedback on academic work.

Challenges associated with the application of AI usage in HE included 'laziness', reduced research synthesis and overall impacts on 'deeper learning', which may lead to a lack of academic progression and reduce student employability. Learning how to develop 'effective prompts' was key for optimizing AI in HE. There was a shared recognition that AI is 'here to stay' and some discussed the importance of integrating AI education into their modules. Going forward, it was felt that equipping both staff and students with the skills to engage with AI critically and effectively, is essential for its responsible use in HE. Further, all groups shared their thoughts on how assessment may need to change to prevent or hinder AI usage and most felt that it may cause the return to formal sit-down examinations to permit a deeper exploration of student learning.

##### 4.2.2.2.1 Sub-theme: Opportunities that develop student learning

Opportunities included personalized learning and mentoring, accessibility and support, yet only with appropriate use, e.g. '*It is a fantastic resource but only if students know how to use it properly'* (TS). AI was described as having the potential to offer tailored learning experiences by adapting content to students' learning speed, strengths and areas for improvement, '*for really personalized learning paths. I think that's the kind of main way I see it working in education'* (TS):

> *"AI provides personalized learning, adaptive learning systems, AI can customize educational content for each student by adjusting to their learning speed, strengths and areas for improvement.....This approach guarantees....a unique and tailored learning experience" (NTS).*

All groups recognized the utility of AI as a mentor '... *almost like a little personal assistant of yours, but also a mentor at the same time'* (SS), with the benefits of having no restrictions on its on usage or time when compared to staff. Aside from the availability of a 24-hour mentor, this mentorship will be personalized to reflect students' learning styles. Peer mentoring was also evident, with students supporting each other in learning how to use AI tools effectively, e.g. '*One of my friends is very positive about AI and encourages me, shows me how to get better outputs'* (NTS).

Furthermore, participants viewed AI as an accessible resource for students who might struggle with traditional academic environments or lack the confidence to speak in class, noting *'It's good for students who are afraid to ask questions in class' (*TS).

AI may be an effective alternative when seeking further information or asking questions. It also assists with accessibility-related challenges, such as breaking down complex information or lengthy texts. Further, the NTS group reflected on AI's ability to streamline aspects of their learning journey, especially close to deadlines. For example, AI models were used as proofreading aids, helping to identify grammatical errors when other services were limited — **'***Proofreading services can only be used so many times in a term… AI fills that gap' (NTS).*

#### 4.2.2.2.2 Sub-theme: Obstacles to effective student learning

Challenges associated with the application of AI usage in HE, stemmed from an overreliance on AI-generated output which they felt led to a lack of academic progression, reduced critical learning skills and impacted on 'deeper learning'. TS voiced concerns that students may mistakenly believe they are making academic progress when instead, they are hindering their learning by reducing their reading and information synthesis skills, e.g. '*I think it's making them very, very lazy and it's making them even lazier, and they think they're doing a good job, which is quite scary'* (TS). This overreliance on AI-generated content may limit students' reading, researching, and synthesizing information skills, which were deemed necessary for not only academic success but also for employability beyond the university context.

Overall, it was expressed that academic integrity, critical thinking and deeper learning may be affected by relying too much on AI output instead of the traditional route of reading, researching, and synthesizing. Overreliance on AI-generated output was a concern discussed across all groups. The NTS group noted that in some cases, their fellow students contributed AI-generated content to assignments without reviewing, fact-checking and editing it, which compromised the significance of the work. For example, a student submitted content to groupwork (presentation) that was irrelevant to the topic:

> *"It was just not correct, you know, and I was maybe gonna have to put this onto a poster and then have to somehow present it, but the information really wasn't about what we were talking about" (NTS).*

SS echoed these concerns as students may incorrectly assume that all AI-generated output is accurate and reliable, e.g. '*this is obviously all legit information that I'm getting….This is perfect, and they can just run with that information'.* The TS group discussed the issues associated with marking student work that included AI-generated content beyond the scope of their course materials. In more technical subjects, such as computer science, students often use AI tools to generate code without evaluating or comparing it to their own attempts. Alongside the consensus that AI has potential to be used as a valuable learning tool in HE, there was a significant challenge in that it could be 'pushing students the other way'.

#### 4.2.2.2.3 Sub-theme: Optimizing AI to enhance education

Participants felt that the key to optimizing AI in HE was to both understand it's limitations and to learn how to develop effective prompts, as they determine the quality and relevance of AI-generated content, e.g. '*I've found it quite good as long as you're specific with the commands you're giving* it' (NTS). This process of refining prompts was likened to a feedback loop - the more tailored the prompt, the more accurate and relevant the output:

> *"I think at the beginning it was kind of just like very vague kind of requests and prompts, but the more you developed them and the more you were being very specific and explicit about what you wanted from the AI model, the better the results were." (SS).*

Interestingly, some felt that being polite when developing prompts could positively influence the model's responses:

> *"If you're nice to the AI model, it gives you better results…kind of like that feedback loop. If you're giving it feedback on how it's performing; it tends to take that on and perform better." (SS).*

Limitations relating to content accuracy were discussed, with comments like *'You definitely cannot just rely to copy and paste and put it back into your document. You have to read through…because it will make mistakes'* (NTS). TS expressed concerns that students may be impressed by the speed and eloquence at which AI can generate content 'very quickly' and that some may assume it is 'gospel'. To overcome this, the need for fact-checking and triangulating content with, for example, peer-reviewed journal articles was emphasized.

One TS exemplified instances where AI omitted the COVID-19 pandemic, which highlights the risk of using outdated or incomplete data. There was consensus that teaching the correct use of AI should be introduced in classroom contexts to support specific elements of assignment planning, such as generating structure or brainstorming ideas. However, the TS and NTS groups were wary of student use particularly when AI was being used for full coursework completion rather than 'scaffolding' of some elements (e.g. structuring work and getting started). The TS and NTS groups suggested that preventing AI in coursework (presentations or group projects) may help maintain a level playing field in student assessment, especially for those who use AI responsibly or choose not to use it at all:

> *"People can show like their talents and, you know, speaking in front of people. And I think it'd be also a great way to ensure that people don't use AI for those. So, like creating a poster and doing a group presentation and a PowerPoint. You know, I feel like you might find less people use AI for those sorts of assignments, because it's just not text" (TS).*

#### 4.2.2.3 Theme 3: The future of AI

Across the focus groups, all participants stated that AI was 'here to stay' and most discussed the need for acceptance of AI, as opposed to resistance. Most participants reflected on how traditional learning models have remained largely unchanged for decades, despite a noticeable shift in student engagement with traditional practices such as reading lists, e.g. '*students are no longer taking reading lists with open arms…you tell me what I need to know to get through this exam'* (SS). These types of students were referred to as the 'quick and easy learners' and a distinction was made between them and those students who are 'deeper learners' who may want to combine both the traditional study methods with the newer forms (AI enhanced learning).

There was a call for institutions to embrace AI proactively and explore how it can support diverse learner profiles — distinguishing between students seeking efficient pathways to qualifications versus those with a deeper academic trajectory, such as postgraduate study. For this to be achieved, all participants would welcome further training to improve their knowledge of AI and increase their interactions and experiences with various AI models. Staff participants were keen to emphasize that they did not want to assume responsibility for further training. Recommendations for training sessions included short and 'easy to digest' sessions which could be run regularly.

At an institutional level, participants considered systemic changes to facilitate and enhance equitable AI use amongst staff and students. For example, one recommendation includes providing access to paid AI tools for all students and staff, to mitigate against digital inequities and ensure fair usage by enabling consistent access and standardized learning experiences. Further institutional level support is needed in the form of well-communicated policies and guidelines alongside the development of AI detection tools. Regarding ethical considerations, participants discussed the need for transparency regarding AI usage - being open and honest about the inclusion of AI content may enhance trust and accountability between students and staff. This openness may reduce concerns about academic integrity and reinforce AI as a learning support tool instead of viewing it as a 'shortcut'. Ethical considerations extended beyond academic honesty as participants also discussed issues around data protection and intellectual property:

> *"I guess we have the wider concerns about the language models themselves, you know, because that's the thing that a lot of text has been scooped up to form the basis of the models which you know has been scooped up without consent. I mean, that's the bigger issue that you're using other people's, the combined work of other people" (TS)*

#### 4.2.2.3.1 Sub-theme: Staff and student training

Participants felt that effective training for both staff and students is essential to support the responsible and effective integration of AI within HE. Input from industry stakeholders would be valuable in developing this training, as it would provide insights into the expectations of employers and help ensure that students are equipped with transferable skills for post-university employment — '*that would be a great idea because we are not currently aware of the extent of AI usage in industry in Northern Ireland*' (TS).

The TS group suggested that training sessions could showcase real-world examples of AI use and offer guidance on current capabilities, whilst also highlighting emerging developments and potential limitations, '*Running rapid fire (training) events on AI usage including examples and ideas and how we might use it and also warning us what's coming down the line as well*' (TS). Short and informative training sessions were called for due to time constraints within academia '*We would need some kind of acceptable use of AI…in a way that's easy to read and easy to digest and doesn't take a week to plough through*' (TS). Further, TS discussed the potential for structured training to be embedded within academic programmes as this would allow students to gain early exposure to effective AI use and have awareness of the advantages and disadvantages of the various AI models.

As discussed under T1 (Attitudes) & T2 (Highlights and Challenges), it was deemed imperative to educate students in the generation of effective prompts and to critically evaluate AI-generated content. This includes teaching awareness that AI often produces 'fluffy' or overly complex language which may require editing to meet academic standards. Future training should emphasise the importance of fact-checking and recognising hallucinations to maintaining academic integrity. Similarly, the NTS group suggested that practical training sessions including these factors would be of benefit, e.g. '*including screenshots or like showing how you've used it. So, like what prompts you've put in? What response you got back...*' (NTS).

AI training for students should also include guidance on research planning and documentation, for example, developing and maintaining records of AI usage to form a transparent academic workflow, ‘So you can see that it's not just copied and pasted it all from ChatGPT, like how you've taken maybe the structure and then where…AI has done the work and where you have done the work and to make it really, really clear’ (NTS). There were also discussions around setting limits on AI usage, such as a proposed 10% cap. However, calculating this with accuracy presents challenges. As such, the focus of future training sessions should remain on promoting responsible use rather than enforcing hard restrictions.

The above themes feature several common themes across the different focus groups. Consistent with the sentiment analysis, some common elements were observed across the different populations. However, key differences also emerged in the focus groups, explored in the following section.

### 4.2.3 NLP Comparison across themes

From the thematic analysis, key themes emerged throughout the focus groups relating to a wide range of topics and experiences. While some common elements were observed across themes in students and staff, these populations also differed in the specific nature of their concerns under each theme. In this section, we explore these similarities and differences between the different participant groups using the NLP pipeline described in Section 3.3.4.

#### 4.2.3.1 Similarities between staff and students

Both students and staff recognize *AI’s utility in academic contexts*, especially for saving time on routine tasks and generating ideas. Participants from both groups gave examples of using tools like ChatGPT to draft emails, brainstorm content, or overcome writer’s block. For instance, a student described using ChatGPT to trim down an overlong assignment paragraph, finding it “quite handy” when they “just couldn’t see where to remove certain unnecessary information”. Similarly, staff members noted AI could effectively *“kickstart”* writing by providing a first draft or ideas, thus helping avoid the “blank sheet”. Both groups mentioned that *good prompts* are key to getting useful output — one staff member and several students learned through experience that *“better prompts mean better output”* and emphasized teaching how to prompt effectively.

Another shared theme is an understanding that *AI is here to stay* in education. Participants acknowledged that outright banning AI is impractical, likening it to *“trying to ban calculators in the 70s”* (as one staff member quipped) and suggesting the need to adapt. A student similarly concluded, *“AI’s here to stay and it needs to be embraced but used safely.”* Both students and staff therefore discussed *educating users on proper AI use* and integrating it with caution. For example, they suggested training students in how to use AI for research (e.g. fact-checking AI outputs, removing AI’s overly “fluffy” or technical language and documenting AI use to maintain transparency. There was broad agreement that *guidance and policy* are needed so that AI becomes a helpful “accessibility tool” without undermining learning or integrity.

*Academic* integrity is a prominent concern for both groups. All participants are aware that AI can be misused (e.g. to cheat or cut corners), and they desire clarity on what is allowed. Both students and staff voiced apprehension about the difficulty of distinguishing AI-generated work and the current lack of reliable detection tools. There is consensus that new strategies are needed to handle this — whether through assessment design, honour codes, or requiring students to show their AI inputs and editing process, aligning with recent research on detection rates (Fleckenstein et al., 2024; Perkins et al., 2024).

In sum, everyone sees AI's double-edged nature: it offers efficiency and support but also poses challenges to academic honesty that universities must navigate.

#### 4.2.3.2 Differences between staff and students

*Enthusiasm vs. Scepticism:* Students generally express more enthusiasm and personal comfort with AI tools, whereas staff tend to be more cautious. Many students view AI as a helpful aid for their work. For example, one student found ChatGPT *"really helpful"* for drafting emails and even a resignation letter, saying it's *"just so much quicker… more creative than I am,"* and that once tweaked, the output *"flows"* better. Several students described positive experiences using AI to generate ideas or simplify tasks. In contrast, staff members were less likely to gush about AI's benefits. While some support staff were very positive (one declared *"I am all for AI. I love using AI"*), many academic staff voiced scepticism about the quality and value of AI-generated content. They frequently pointed out that ChatGPT's output can be *"pure garbage"* or overly generic, requiring an expert eye to filter or improve it. One faculty participant described moving away from using AI because *"the text it generates is so bland… it's pointless."* This difference in tone suggests that *students tend to focus on AI's conveniences, while staff emphasize its shortcomings* and the effort needed to correct or vet AI output.

*Academic Integrity & Cheating:* Both groups worry about cheating, but the *focus of their concern differs.* Staff members are concerned system-wide — they cited cases of students using ChatGPT to generate *admissions personal statements* or exam answers, which means the work is not the student's own. Staff lamented that *"nobody's writing personal statements, they're all going to ChatGPT"*, and described receiving AI-written external examiner applications and student assignments that give them *"no sense"* of the real person's ability. This *erodes trust* in submitted work for educators. Some staff even discussed reverting to in-person exams and supervised assessments, feeling that *"AI is definitely going to force us back"* to more controlled forms of testing to ensure authenticity. Students, on the other hand, voiced concerns on a more personal level — they worry about *fairness and the risk of false accusations.* Students don't want to be unfairly suspected of cheating if they use AI appropriately. One student noted fear that *using AI even once could get them flagged* by plagiarism detectors and "scare them off" using it again. Another student, while initially *"worried it [using ChatGPT] was a bit cheating,"* ultimately used it for minor help (like trimming words) and found that acceptable. In summary, *staff focus on preventing dishonest use,* even if it means stricter assessment rules, whereas *students focus on being allowed to use AI legitimately without undue suspicion.*

*Learning and Assessment:* Students and staff also diverge in how they think AI affects learning. A number of *students acknowledged the learning drawbacks if* AI is overused. For example, one student pointed out that in a course with no exams, *"if somebody is using [AI] for all of their coursework, you don't know if they've actually learned anything or if it's just all coming from ChatGPT."* This shows students' awareness that *relying on AI can undermine genuine learning.* Staff share this concern but approach it from a teaching standpoint: some faculty discussed redesigning assessments to be "authentic" and less vulnerable to AI misuse (such as oral presentations or iterative in-class work). There was internal debate among staff — some felt they *"would like more tests… just so we can more accurately assess ability,"* while others noted the push for open-ended authentic assessments that don't easily allow a single AI-generated answer. Students did not delve deeply into assessment policy, but a few offered ideas like requiring students to submit *AI usage logs or reflections* with their work to prove their learning process. Generally, *staff view AI as a challenge to traditional assessment validity,* whereas *students view it as a tool that should be integrated with accountability measures* so that they can use it without devaluing their work.

*Use Cases:* The types of AI use described also highlight differences. Students mainly discussed *using AI for personal productivity* — e.g. improving writing style, generating examples, summarizing or editing text, and getting unstuck on assignments. Staff, beyond using AI for their own workflow (like lesson prep or administrative tasks), also talked about *observing others' use* (students, colleagues) and its institutional implications. For instance, support staff talked about AI in marketing or design tasks (e.g. image editing, press release drafting), and academic staff discussed AI use by partner colleges and students in various processes (revalidation documents, applications, etc.) that they oversee. In short, *students focus on how AI helps them as learners,* whereas *staff also consider broader impacts on educational operations.*

*Trust in AI Outputs: staff* exhibit lower trust in the accuracy and originality of AI-generated content. Some staff mentioned reading about bias and factual errors in AI outputs, and a general sentiment was that AI text might be *"well-written"* yet not truly reliable or deep. By contrast, students did not explicitly discuss hallucinations or factual inaccuracies as much, perhaps because they use AI for relatively straightforward tasks. However, a few students did notice issues — for example, a psychology student found that ChatGPT provided non-scholarly sources (*"simply psychology"* websites) as references, which *"is just a no"* in academic work. Overall, though, *students seemed more ready to accept AI outputs at face value* for things like grammar or wording, whereas *staff were warier* and insisted on fact-checking and editing any AI-produced material heavily.

## 5 Discussion

In this study, we investigated students' and staff perceptions of AI in education contexts, across three different qualitative methodologies — free text comments, Padlet comments, and focus groups — using four qualitative analytic approaches: sentiment analysis, word cloud analysis, stance analysis, and a thematic analysis complemented with an NLP thematic comparison. In this section, we discuss the variation between these analyses, both in general and in the context of previous studies.

### 5.1 Analyses in the current study

First, as seen in Section 4.1, the mean sentiment scores for both students and staff are generally close to the midpoint of 0, relative to the range of -1 to 1. However, these averages should not be interpreted as representing overall neutrality; neutral responses represent only 44.5% of responses for the survey data and 32.1% of the responses for the Padlet data, with the majority being clearly positive or negative. For both students and staff, these scores present AI as a polarizing issue, in which a growing awareness of the pros and cons of AI use has yet to lead to a balanced consensus.

For the sentiment analysis, no significant differences were observed between student and staff sentiment scores. However, this wide variation within both groups is reflected throughout the remaining analyses. For example, the word cloud analysis included content relating to various aspects of AI in education contexts — suggesting a wide range of experiences in relation to AI. Similarly, the thematic analysis identified several common themes across the population groups — including both positive and negative perceptions. Crucially however, the comparison across themes did identify contrasting perceptions between students and staff towards AI in education contexts — in contrast with the sentiment analysis.

This contrast highlights the variation that can be attributed to different approaches to investigating student and staff perceptions. That is, the nature of the perceptions identified through the sentiment analysis vs through thematic analysis can be related to the methodologies used to elicit them. In general, the survey comments tend to be the most cursory; as described above, these were entered into a free-form text field by participants who had already completed a survey asking more explicit questions about different aspects of their familiarity with AI. However, as Example (3) shows (cf. Section 4.1.1), some participants still used this comment field as an opportunity for more detailed discussion.

The Padlet platform provided the opportunity for more extensive discussion, and allowed participants to create threads in which they responded to each other's comments; however, some of these discussions developed in ways that did not directly address the issues mentioned in the original Padlet prompt, and the unmoderated nature of the discussions provided no opportunity for a more detailed exploration of issues raised, unless such an exploration was initiated spontaneously by participants. The perceptions identified through these media thus provide a valuable indication of which topics struck the participants as most immediately relevant; however, they also display more heterogeneity and less depth than could be obtained through the focus groups.

With these considerations, the research question of perceptions in education contexts does not have a single answer: whether students and staff perceptions vary depends on the measure of perceptions. This caveat has key implications for practical applications that draw on empirical evidence from perceptions studies. For example, for institutional guidance on AI that relates to a specific educational domain (e.g. staff or student training), respective evidence for developing the guidance should include perceptions relating to the same specific domain. In contrast, overall measures like sentiment scores can be useful for comparing with other individual difference factors, e.g. demographics or subject area, to identify broader trends in perceptions (Gerard et al., 2025; e.g. Petricini et al., 2024).

## 5.2 Connection to Previous Literature

This study aligns with and extends existing literature on perceptions of AI in higher education, particularly in terms of both positive and negative themes. Across all data types (i.e. free-form survey comments, Padlet discussions and focus groups) participants expressed views consistent with those reported in prior single-method studies (Chan & Hu, 2023; Salas-Pilco et al., 2023; Zhou et al., 2024).

### 5.2.1 Positive Themes and Methodological Alignment

Across multiple sources, participants (especially students) frequently articulated positive attitudes towards AI as a tool for improving efficiency, creativity, and access. This supports findings from Chan and Hu (2023) and Zhou et al. (2024), who identified practical benefits such as assistance with brainstorming, summarising and improving written content. In our data, this was especially evident in free-form and Padlet responses, which, due to their asynchronous and reflective nature, highlighted AI's role in overcoming initial writing barriers and synthesising information under time constraints. Focus group participants elaborated further on the affordances of AI for neurodivergent users, describing it as a supportive 'mentor' or 'personal assistant', a theme less visible in earlier work using narrower survey instruments.

Moreover, staff and non-teaching participants frequently described AI as a valuable administrative tool, echoing findings from Zawacki-Richter et al. (2019) and Salas-Pilco et al. (2023). These benefits emerged most clearly in Padlet posts and focus group transcripts, where discourse-pragmatic analysis captured expressions of enthusiasm, curiosity, and professional adaptability.

### 5.2.2 Negative Themes and Methodological Alignment

Conversely, our findings also affirm and extend concerns previously raised about the risks of AI in educational settings. Staff and student concerns about academic integrity, depersonalisation, and skill erosion were consistent with earlier work (Zhou et al., 2024; Zawacki-Richter et al., 2019). However, our use of multi-method qualitative design allowed us to unpack these concerns in greater detail. Focus groups in particular enabled the emergence of richer, dialogic expressions of ethical and pedagogical unease, including fears of 'cheating', loss of student identity and the weakening of core academic competencies such as reading, synthesis, and critical reflection.

In contrast to Padlet and free-text survey comments, which tended to foreground task-specific positives, focus group discussions often revealed a deeper layer of ambivalence or outright opposition to uncritical AI adoption. This supports the view that methodological context (e.g., synchronous dialogue versus asynchronous writing) significantly affects the depth and emotional tone of participant perceptions (Vu et al., 2025). Indeed, the combination of thematic, sentiment, and discourse-pragmatic analysis revealed that participants often shifted between enthusiasm and concern depending on the communicative setting and prompt.

## 5.3 Conclusion

By integrating diverse qualitative methods, this study supports and complicates existing findings in the literature. While it corroborates widely reported positive themes around efficiency, creativity, and administrative use, it also reveals how emotional tone, stance, and rhetorical function vary with method. Our triangulated approach clearly demonstrates the need to move beyond single-method studies and engage with the discursive construction of AI perceptions, capturing not only what participants think but how and why they express it.

# 6 Conflict of Interest

The authors declare that the research was conducted in the absence of any commercial or financial relationships that could be construed as a potential conflict of interest.

# 7 Author Contributions

Conceptualisation, J.G., K.N, M.S.; methodology, J.G., M.M., K.N., M.S.; software, M.M.; validation, J.G., M.M., K.N.; formal analysis, M.M., K.N., M.S.; investigation, J.G., M.M. K.N., M.S.; data curation, M.M., K.N., M.S.; writing - original draft, M.M. K.N., A.R., M.S.; writing - review and editing, J.G.; visualization, M.M., M.S.; supervision, J.G.; project administration, J.G.; funding acquisition, J.G., K.N., A.R, M.S. All authors have read and agreed to the published version of the manuscript.

# 8 Funding

This study was funded by an award from the International Science Partnerships Fund (ISPF).

# 9 Acknowledgments

We are thankful to the ISPF project team for support in developing the questionnaire, including Muhammad Usman Hadi (Ulster University), Antoine Rivoire (Ulster University), Zhiwei Lin (Queen's University Belfast), Jocelyn Dautel (Queen's University Belfast), Adina Camelia Bleotu (University of Bucharest) and Caitlin Meyer (University of Amsterdam), as well as Louise Drumm (Edinburgh Napier University) for helpful discussion on the Padlet comments, and Sahajpreet Singh for contribution to the sentiment scores analysis. We would also like to thank Sue Attewell (JISC) and Nicola Bartholomew (Ulster University), as well as the attendees of the workshop on Adaptive Education: Harnessing AI for Academic Progress for helpful discussion. Last but not least, we greatly appreciate the time spent by the survey participants from Ulster University.

## 12 Data Availability Statement

The datasets generated are available upon request from the authors.

**Table A1. Mean sentiment scores — both RoBERTa model and manual scoring on -1 to +1 scale.**

| Data source | Population | RoBERTa model score | Manual sentiment score |
|---|---|---|---|
| Free-form text comments | Undergraduate | -0.02 (SD = 0.744) | 0.05 (SD = 0.757) |
| | Postgraduate | -0.27 (SD = 0.604) | -0.19 (SD = 0.634) |
| | Teaching staff | -0.23 (SD = 0.762) | -0.16 (SD = 0.779) |
| | Non-teaching staff | 0.06 (SD = 0.712) | 0.00 (SD = 0.728) |
| Padlet comments | Students | -0.10 (SD = 0.764) | 0.03 (SD = 0.814) |
| | Staff | -0.09 (SD = 0.812) | 0.17 (SD = 0.851) |